\documentclass{aa}
\DeclareUnicodeCharacter{2061}{}
\usepackage[varg]{txfonts}
\usepackage{amsmath}
\usepackage{dblfloatfix}
\usepackage{placeins}
\usepackage{natbib}
\usepackage{graphicx}
\usepackage[breaklinks=true]{hyperref}
\bibpunct{(}{)}{;}{a}{}{,}

\begin{document}

   \title{Suppression of water-ice UV photodesorption on cosmic dust analogs}

   \subtitle{}

   \author{Ph. Parent\inst{1}\corrauth{philippe.parent@univ-amu.fr}
        \and C. Laffon\inst{1}\email{carine.laffon@univ-amu.fr}
        }

   \institute{Aix Marseille Univ, CNRS, CINaM, Marseille, France}

   \renewcommand{\today}{November 6, 2026}
   \date{}

\abstract
{Ultraviolet (UV) photodesorption regulates the partitioning of water between interstellar ice mantles and the gas phase, yet laboratory yields are generally measured on planar substrates rather than on surfaces representative of cosmic dust. We investigate how grain morphology affects UV photodesorption from thin H$_2$O ice films of 2--6 monolayers deposited on planar gold and olivine, soot nanoparticles, and olivine microparticles. Samples were irradiated at 12 and 60 K with a microwave-discharge hydrogen lamp, and ice depletion was monitored by X-ray photoelectron spectroscopy and reflection-absorption infrared spectroscopy. On planar gold, the photodesorption yield increases approximately linearly with ice coverage, from $\sim5\times10^{-4}$ to $\sim2\times10^{-3}$ molecule photon$^{-1}$ between 2 and 6 monolayers, consistent with increasing UV absorption in optically thin ice; planar olivine gives comparable yields. In contrast, photodesorption is suppressed on particulate olivine and, more strongly, on soot, where yields are up to a factor of $\sim40$ lower than on gold at 12 K. Both experimental techniques confirm the strong suppression of ice depletion on soot. These results show that UV photodesorption depends not only on ice coverage but also strongly on grain morphology. The pronounced suppression on soot may reflect heterogeneous ice growth and altered energy redistribution within nanoscale ice domains, implying that yields measured on planar substrates may substantially overestimate photodesorption from the thin ice mantles of cosmic dust.}
   \maketitle
   \nolinenumbers

\section{Introduction}

Although gas-phase chemistry contributes to H$_2$O formation in diffuse and translucent clouds, in cold, dense interstellar and circumstellar environments water is largely formed through grain-surface hydrogenation reactions, and the icy mantles of cosmic grains constitute its major reservoir \citep{Boogert2015,VanDishoeck2013,Oberg2011}. At the low dust temperatures prevailing in molecular clouds, prestellar cores, and outer protostellar envelopes, thermal sublimation of water is negligible, and ultraviolet (UV) photodesorption is an important process releasing H$_2$O from water ice into the gas phase, along with UV photoproducts such as H$_2$, OH, and O$_2$ \citep{Cruz-Diaz2018}.

The importance of UV photodesorption, however, changes with depth into the cloud as the UV radiation field becomes attenuated. Astrochemical models indeed predict a stratified cloud structure in which the gas--ice partitioning of water results from the competition between ice formation, UV photodesorption, photodissociation, and freeze-out. Close to the cloud edge, UV radiation prevents ice accumulation. At intermediate extinctions, increasing shielding allows ice mantles to grow, while residual UV photodesorption continues to return H$_2$O to the gas phase; deeper inside the cloud, a residual gas-phase abundance is sustained by cosmic-ray-induced UV photons \citep{Hollenbach2009}, with sputtering providing an additional contribution \citep{Silsbee2021}. Herschel observations of prestellar cores and cold protostellar envelopes support this stratified picture, revealing enhanced gaseous H$_2$O in externally irradiated outer layers and a much lower abundance deeper inside \citep{VanDishoeck2021}. UV photodesorption can also influence the gas--ice partitioning of water in the cold outer regions of protoplanetary disks \citep{Krijt2016}. The gas-phase water abundance in astrophysical environments is therefore sensitive to the UV photodesorption yield, making measurements under realistic grain and ice conditions essential.

As discussed below, UV photodesorption is a surface process, but its
efficiency depends on photon absorption and energy redistribution within
the upper molecular layers of the ice. The thickness of interstellar ice
mantles is therefore an important parameter. \citet{Boogert2015} compiled H$_2$O ice column densities measured along quiescent lines of sight through nearby molecular clouds and dense cores. Such column densities cannot be converted uniquely into the thickness of an individual grain mantle because the number and size distribution of the ice-bearing grains intercepted along the line of sight are not known. As detailed in Appendix~\ref{app:coverage}, assuming a Mathis--Rumpl--Nordsieck (MRN) grain-size distribution, the total grain surface area is $A_{\rm surf/H}\simeq6.1\times10^{-21}$~cm$^{2}$~H$^{-1}$, such that one complete monolayer corresponds to an H$_2$O abundance of $\simeq6.1\times10^{-6}$ per H nucleus, which translates into typical effective H$_2$O coverages of about 5--8~ monolayers (ML). Including other abundant ice constituents, in particular CO and CO$_2$, would make the total ice coverage $\sim20$–30\% larger (6--10~ML) \citep{Boogert2015}. Furthermore, the budget of available oxygen included in H$_2$O gives upper limits to the mantle thicknesses to approximately 50~\AA\ ($\sim$15~ML) \citep{Boogert2015}. These values are substantially below the thick ice mantles commonly invoked in the astrochemical literature, from $\sim$100~ML in dense-cloud models (e.g. \citealt{Cuppen2017}) to coverages as high as 12,000~ML obtained by simply dividing the observed column density by the surface density of one monolayer, without accounting for the total grain surface area intercepted along the line of sight \citep{Potapov2020}. The 2--10 ML range investigated here is therefore representative of
interstellar ice mantles, while remaining below the practical upper
thickness limit of both X-ray photoelectron spectroscopy (XPS) and
reflection--absorption infrared spectroscopy (RAIRS). Previous experimental studies have shown that, at such low coverages, surface morphology strongly affects molecular adsorption, sticking, and the growth of water ice on realistic grain analogs \citep{Laffon2021}. It may likewise influence the efficiency of UV photodesorption from the resulting ice. This provides a direct motivation for investigating photodesorption from grain analogs and comparing it with the planar surfaces conventionally used to determine UV photodesorption yields for astrochemical modeling.

UV photodesorption of water ice was first quantified by \citet{Westley1995}, who measured the photoinduced mass loss of thick films and obtained UV yields increasing from about $3.5\times10^{-3}$ at 40~K to $8\times10^{-3}$ molecule~photon$^{-1}$ at 100~K. Molecular-dynamics simulations identified several microscopic photodesorption pathways, including direct escape of H and OH, recombination-induced H$_2$O desorption, and kick-out of a neighboring water molecule \citep{Andersson2008}, and predicted combined OH + H$_2$O photodesorption yields of about $(5$--$7)\times10^{-4}$ molecule~photon$^{-1}$ between 10 and 90~K \citep{Arasa2010}. RAIRS measurements by \citet{Oberg2009} found UV photodesorption yields of a few $10^{-3}$ molecule~photon$^{-1}$, increasing with both temperature and ice thickness. \citet{Cruz-Diaz2018} subsequently measured the gas-phase products separately, detecting H$_2$O, H$_2$, OH, and O$_2$. The intact H$_2$O yield increased from approximately $1.5\times10^{-3}$ at low temperature to $3\times10^{-3}$ molecule~photon$^{-1}$ around 70--90~K, whereas the combined H$_2$O + OH + O$_2$ yield reached about $7\times10^{-3}$ molecule~photon$^{-1}$. Energy-resolved experiments further demonstrated a dependence on the UV spectrum, isotopic composition, and ice morphology \citep{Fillion2022}, and \citet{Bulak2023} showed that solid-state photolysis may deplete molecular H$_2$O more rapidly than desorption. Overall, photodesorption yields of water ice are typically of the order of a few $10^{-3}$ molecule~photon$^{-1}$, but vary significantly with temperature, ice thickness, irradiation spectrum, and ice morphology.

On planar substrates, the initial ice column density and its evolution under irradiation can be determined easily, providing quantitative UV photodesorption yields. On analogs of cosmic grains, such measurements are more difficult because converting an experimental signal into an ice thickness generally requires knowledge of the accessible microscopic surface area of the particulate substrate, which is poorly defined for porous or aggregated materials. A recently developed approach based on XPS circumvents this limitation \citep{Laffon2021, Stadler2024, Parent2026}. Substrate-specific uptake curves allow the XPS signal to be converted into an equivalent coverage without independently determining the grain surface area. Photodesorption yields can therefore be quantified with XPS on particulate substrates as well as on planar references, where RAIRS can also be used \citep{Laffon2025}.

Here, we combine XPS and RAIRS to investigate the UV processing of thin water-ice films deposited on four substrates: planar gold, planar olivine, soot nanoparticles, and olivine microparticles, the latter two serving as carbonaceous and silicate grain analogs, respectively. This set of substrates enables a direct comparison between planar and particulate morphologies and between chemically distinct carbonaceous and silicate materials.

\section{Methods}
\label{sec:methods}

\subsection{Experimental setup and substrates}
\label{sec:substrates}

The experiments were performed in the SUMO ultrahigh-vacuum setup, which combines UV irradiation with in situ XPS and RAIRS \citep{Laffon2025}. The samples were mounted on the cold finger of a closed-cycle cryostat cooled by a SHI Cryogenics DE204B cryocooler, providing a minimum temperature of $12\pm1$~K. The temperature was measured using a Lakeshore DT-670 silicon diode positioned close to the sample and regulated by a proportional--integral--derivative (PID)-controlled 50~W resistive heater.

Four substrates were investigated. The planar reference was a $25\times25$~mm$^2$ gold foil (99.99\%, Alfa Aesar). A cut and polished $10\times10\times1$~mm$^3$ peridot crystal from San Carlos was used as a planar olivine substrate. A particulate olivine sample was prepared by drop-casting an ethanolic suspension of finely ground San Carlos olivine onto a support mounted on the same sample holder as the planar crystal. Fine grinding produced particles with a mean equivalent diameter of $0.30\pm0.14$~$\mu$m, together with a minor population of larger grains several micrometers in size, as determined by scanning electron microscopy (SEM; JEOL 7900-F) (see Sect.~\ref{sec:results}). Carbon soot was produced from the flame of a pure paraffin candle and deposited directly onto half the surface of the gold foil. It consists of fractal aggregates formed from primary nanoparticles of turbostratic graphite with a mean diameter of $25\pm15$~nm \citep{Laffon2021}. The morphology of the soot particles was characterized by transmission electron microscopy (TEM) using a JEOL JEM-2010 operated at 200~kV in bright-field mode. An effective nanoscale roughness of the soot surface was estimated from TEM images using cross-sectional line profiles. This analysis yielded characteristic depths $d \simeq0.5$~nm and widths $w \simeq2$~nm, with the projected surface approximately equally divided between concave and convex regions \citep{Parent2026}. The soot film was additionally imaged by SEM using the same JEOL 9700-F microscope.

\subsection{UV irradiation and spectroscopic measurements}
\label{sec:spectroscopy}

UV irradiation was provided by a home-built T-type microwave-discharge hydrogen lamp (MDHL) equipped with a MgF$_2$ window, coupled to an Opthos radio-frequency (RF) generator and resonant cavity. The lamp emits H Ly$\alpha$ at 121.6~nm together with molecular H$_2$ bands between 110 and 180~nm. The photon flux at the sample was calibrated with an extreme-ultraviolet (XUV) photodiode and continuously monitored during irradiation. Irradiation was typically performed at $3\times10^{13}$~photons~cm$^{-2}$~s$^{-1}$ for 800~min, corresponding to a fluence of $1.5\times10^{18}$~photons~cm$^{-2}$. Reflection--absorption infrared spectra were recorded with a JASCO FT/IR-6300V spectrometer using a liquid-nitrogen-cooled mercury cadmium telluride (MCT) detector. Reflection occurred at a grazing angle of $6\pm2^\circ$ relative to the surface plane. Spectra were obtained by averaging approximately 700 scans at a resolution of 1~cm$^{-1}$.

XPS measurements were performed using a non-monochromatized PSP Vacuum TX400 dual-anode X-ray source and a PSP Vacuum Resolve 120 hemispherical analyzer. The Mg K$\alpha$ line at 1253.6~eV was used with an anode power of 100~W. The source-to-sample distance was adjusted to minimize X-ray-induced photolysis or photodesorption. The absence of measurable X-ray-induced modification was verified by monitoring the stability of the RAIRS signal from a water ice film during XPS acquisition in the absence of UV irradiation. Photoelectrons were collected at a take-off angle of $55^\circ$ relative to the surface normal. At this geometry, the mean path length through an overlayer coating spherical particles---as in the case of carbon soot particles covered with ice---is equivalent to that through an overlayer on a planar surface. This ``magic-angle'' geometry enables comparison of overlayer signals measured on particulate and flat substrates \citep{Gunter1997,Gunter1995,Kappen2000}, as previously verified for water condensed on soot films similar to those used here \citep{Laffon2021}.

\subsection{Ice deposition and coverage calibration}
\label{sec:coverage}

Water vapor was produced from deionized liquid water purified by repeated freeze-pump-thaw cycles and introduced isotropically into a dedicated dosing chamber through a precision leak valve. The incident molecular flux was estimated from the background pressure using the Hertz--Knudsen equation. Exposures are expressed in Langmuir, with 1~L $=10^{-6}$~Torr~s. Under the present deposition conditions, the H$_2$O sticking coefficient on gold was close to unity, such that approximately one monolayer was condensed per Langmuir of exposure. This was verified with RAIRS from the column density derived from the integrated OH-stretching band centered near 3257~cm$^{-1}$, using a band strength of $A=2.0\times10^{-16}$~cm~molecule$^{-1}$ \citep{Bouilloud2015}. The integrated absorbance increased linearly with exposure up to approximately 10~ML for films deposited at 12 and 60~K. At larger thicknesses, optical interference causes deviations from
linearity \citep{Cholette2009}, preventing quantitative RAIRS coverage
determination.

As detailed in Appendix~\ref{app:xps}, XPS uptake curves for each substrate, $R_{\rm XPS}$, were obtained for gold, soot, and olivine by normalizing the H$_2$O O~1s signal to the corresponding substrate signal \citep{Laffon2021}. These curves were used to determine the H$_2$O sticking coefficients on the different substrates. Relative to $S\simeq1$ on gold, the sticking coefficients were approximately 0.2 on soot, 0.5 on olivine particles, and 0.8 on the olivine crystal. The uptake curves were also fitted and analytically inverted to provide a calibration relating $R_{\rm XPS}$ to coverage. Because the XPS response becomes progressively less sensitive to additional ice as the film thickness approaches the photoelectron attenuation length (Fig.~\ref{fig:xps_calibration}), we have restricted our XPS measurements to initial coverages of about 6~ML.

During UV irradiation, the XPS and RAIRS signals were recorded simultaneously as a function of irradiation time. For XPS, each measured $R_{\rm XPS}$ value was converted into the corresponding ice coverage using the inverse of the uptake curve measured on the same substrate. For RAIRS measurements on gold, the integrated OH-stretching intensity was directly converted into ice coverage. On soot, the RAIRS signal was scaled to the initial
coverage estimated from the H$_2$O exposure and the XPS-derived
sticking coefficient, without requiring knowledge of the substrate
surface area. On olivine crystal and olivine microparticulate substrates, the RAIRS response is strongly affected by the substrate's absorption, preventing quantitative use of the band intensities. These procedures yield the XPS- and RAIRS-derived coverage curves presented in Fig.~\ref{fig:gold_soot}. Irradiation time was converted into photon fluence using the measured UV photon flux, and the slopes of the resulting coverage-versus-fluence curves were used to determine the corresponding yields, as defined below.

\subsection{Definition of XPS- and RAIRS-derived photodesorption yields}
\label{sec:yields}

UV processing of water ice may produce three distinguishable effects: photodesorption, which removes H$_2$O molecules or their fragments from the condensed phase; molecular photoconversion, which transforms H$_2$O into photoproducts remaining in the ice, such as OH, O$_2$, O, HO$_2$, and H$_2$O$_2$ \citep{Laffon2006}; and structural photoconversion, which modifies the hydrogen-bonded network and its infrared response. These processes affect the XPS and RAIRS signals differently. The XPS O1s signal is dominated by H$_2$O, since the concentrations of trapped oxygen-bearing photoproducts in irradiated ice are expected to be of order 1\% or less \citep{Bulak2022,Laffon2006}. A decrease in the XPS coverage is therefore taken as a measure of the net loss of material from the water ice and expressed as a photodesorption yield, $Y_{\rm XPS}$. In contrast, the RAIRS OH-stretching intensity is affected by
photodesorption, molecular photoconversion, and UV-induced structural
changes, so that the resulting photodepletion yield,
\(Y_{\rm RAIRS}\), includes these three contributions \citep{Bulak2023}.

\section{Results}
\label{sec:results}

\subsection{UV irradiation of water ice on gold and soot}
\label{sec:gold_soot}

Figure~\ref{fig:gold_soot}A,B compares the evolution of the XPS- and
RAIRS-derived coverages during UV irradiation of ice films deposited on
gold and soot. Initial coverages of approximately 2, 4, and 6~ML were
deposited and irradiated at 12~K (solid lines) and 60~K (dashed lines).

On the gold substrate, both the XPS- and RAIRS-derived coverages
decrease continuously with photon fluence for all initial coverages and
at both temperatures, with a more pronounced decrease in the
RAIRS signal. On soot, by contrast, both the XPS and RAIRS signals
remain nearly constant, indicating only weak depletion. Thus, both
techniques consistently show that UV-induced ice depletion is strongly
suppressed on soot.

\subsection{Determination of the yields}
\label{sec:yield_determination}

For both techniques, the yield $Y$, expressed in
molecule~photon$^{-1}$, was derived from the decrease in coverage,
$\theta$, as a function of incident photon fluence, $\Phi$. Taking one
monolayer to correspond to
$N_{\rm ML}=10^{15}$~molecules~cm$^{-2}$, the yield is

\begin{equation}
Y=-N_{\rm ML}\frac{{\rm d}\theta}{{\rm d}\Phi}.
\label{eq:yield}
\end{equation}

\begin{figure*}[t]
\centering
\includegraphics[width=13cm]{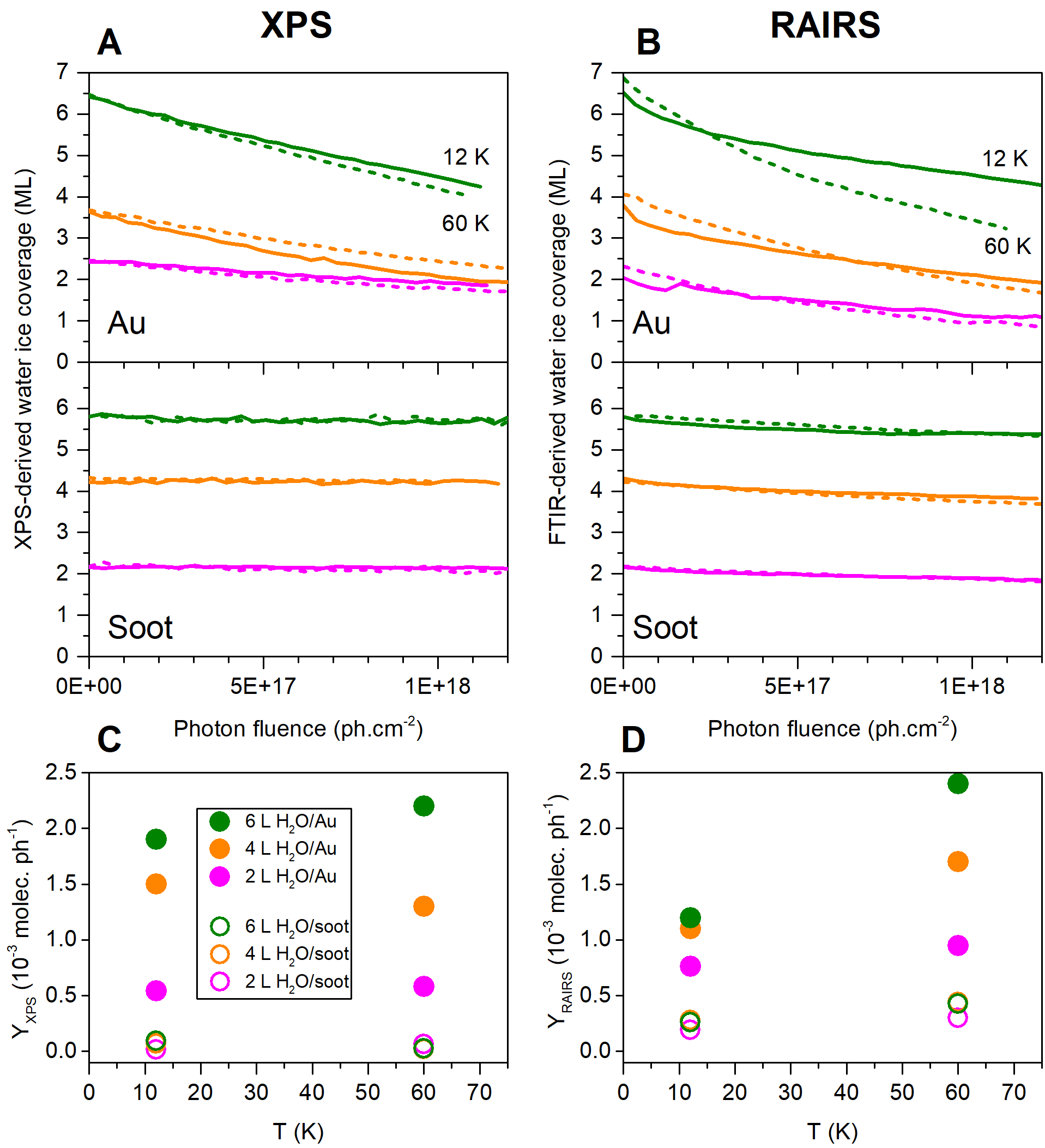}
\caption{Comparison of the XPS- and RAIRS-derived evolution of water
ice during UV irradiation on gold and soot.
(A) Coverage derived from XPS measurements.
(B) Coverage derived from RAIRS. Porous ice films with different
initial coverages were deposited and irradiated at 12~K (solid lines)
and 60~K (dashed lines).
(C) Photodesorption yields $Y_{\rm XPS}$ obtained from the slopes of
the XPS-derived coverages.
(D) Photodepletion yields $Y_{\rm RAIRS}$ (i.e. photodesorption plus
molecular and structural photoconversion) obtained from the slopes of
the linear regions of the RAIRS-derived coverages.
Filled symbols represent water ice deposited on gold, whereas open
symbols represent ice deposited on soot. Magenta, orange, and green
symbols correspond to nominal initial coverages of 2, 4, and 6~ML,
respectively. Numerical yield values are listed in
Table~\ref{tab:yields}. The estimated relative uncertainties are
$\pm25$\% for $Y_{\rm XPS}$ and $\pm30$\% for $Y_{\rm RAIRS}$
(see Appendix~\ref{app:xps}).}
\label{fig:gold_soot}
\end{figure*}

Linear fits over the full fluence range were used to derive the XPS
photodesorption yields, \(Y_{\rm XPS}\). The RAIRS depletion curves
(Fig.~\ref{fig:gold_soot}B) show an initial nonlinear decrease,
attributed to photolysis and UV-induced structural changes, followed by
a linear regime mainly associated with photolysis, as proposed by
\citet{Oberg2009}. The RAIRS photodepletion yields,
\(Y_{\rm RAIRS}\), were therefore derived from this linear regime only,
starting at a fluence of approximately
\(2\times10^{17}\)~photons~cm\(^{-2}\).
The resulting \(Y_{\rm XPS}\) and \(Y_{\rm RAIRS}\) yields are shown in
Fig.~\ref{fig:gold_soot}C,D, respectively, and listed in
Table~\ref{tab:yields}. On gold, both yields increase with initial ice
coverage, from approximately \(5\times10^{-4}\) molecule photon\(^{-1}\)
for 2~ML to about \(2\times10^{-3}\) molecule photon\(^{-1}\) for
6~ML. Only limited variations are observed between 12 and 60~K. The
\(Y_{\rm RAIRS}\) values are comparable in magnitude to the
\(Y_{\rm XPS}\) values. The \(Y_{\rm XPS}\) values are of the same
order as the photodesorption yields of intact H$_2$O reported by
\citet{Cruz-Diaz2018}, who obtained
\((1.3\pm0.2)\times10^{-3}\) molecule photon\(^{-1}\) at 8~K and
\((1.6\pm0.3)\times10^{-3}\) molecule photon\(^{-1}\) at 50~K.
Likewise, the \(Y_{\rm RAIRS}\) values are consistent with the RAIRS
measurements of \citet{Oberg2009}, who reported yields of a few
\(10^{-3}\) molecule photon\(^{-1}\) with systematic dependences on
temperature and ice thickness.

The photodesorption yield per incident photon depends
on the fraction of the incident UV radiation absorbed by the ice. In
the optically thin limit, the Beer--Lambert law gives

\[
P_{\rm abs}\simeq \sigma N_{\rm ML}\theta ,
\]

where \(\sigma \simeq 3.4\times10^{-18}\) cm\(^{2}\) is the average
VUV absorption cross section of solid H$_2$O over the 120--165 nm
range \citep{CruzDiaz2014}, \(N_{\rm ML}=10^{15}\) molecules
cm\(^{-2}\), and \(\theta\) is the ice coverage in ML. Even at
6~ML, the absorbed fraction is only about 2\%, confirming that the ice
remains optically thin over the investigated coverage range. The
fraction of absorbed photons therefore increases approximately
linearly with ice coverage.
On gold, for the three measurements at 12~K, the mean value of
\(Y_{\rm XPS}/\theta\) is
\(3.2\times10^{-4}\) molecule photon\(^{-1}\) ML\(^{-1}\), so that
the measured coverage dependence can be expressed as

\begin{equation}
Y_{\rm XPS}^{\rm Au}(\theta)
\simeq 3.2\times10^{-4}\,\theta
\quad {\rm molecule~photon^{-1}},
\label{eq:yield_au}
\end{equation}

\begin{table}[t]
\caption{UV photodesorption and photodepletion yields
(molecule~photon$^{-1}$) measured by XPS and RAIRS, respectively.}
\label{tab:yields}
\centering
\resizebox{\columnwidth}{!}{
\begin{tabular}{@{}lcccc@{}}
\hline\hline
 & XPS 12~K & XPS 60~K & RAIRS 12~K & RAIRS 60~K \\
\hline
\multicolumn{5}{c}{Au} \\
2~ML &
$5.4\times10^{-4}$ &
$5.8\times10^{-4}$ &
$7.6\times10^{-4}$ &
$9.5\times10^{-4}$ \\

4~ML &
$1.5\times10^{-3}$ &
$1.3\times10^{-3}$ &
$1.1\times10^{-3}$ &
$1.7\times10^{-3}$ \\

6~ML &
$1.9\times10^{-3}$ &
$2.2\times10^{-3}$ &
$1.2\times10^{-3}$ &
$2.4\times10^{-3}$ \\
\hline

\multicolumn{5}{c}{Soot} \\
2~ML &
$1.4\times10^{-5}$ &
$1.3\times10^{-4}$ &
$1.9\times10^{-4}$ &
$3.0\times10^{-4}$ \\
&
$4.6\times10^{-5}$ (*) &
&
&
\\

4~ML &
$7.2\times10^{-5}$ &
$1.9\times10^{-5}$ &
$2.8\times10^{-4}$ &
$4.4\times10^{-4}$ \\

6~ML &
$9.5\times10^{-5}$ &
$2.7\times10^{-5}$ &
$2.6\times10^{-4}$ &
$4.3\times10^{-4}$ \\

10~ML &
-- &
-- &
$3.4\times10^{-4}$ ($\dagger$) &
-- \\
\hline

\multicolumn{5}{c}{Olivine crystal} \\
4~ML &
$1.8\times10^{-3}$ &
-- &
-- &
-- \\
\hline

\multicolumn{5}{c}{Olivine particles} \\
4~ML &
$7.2\times10^{-4}$ &
-- &
-- &
-- \\
\hline
\end{tabular}
}
\tablefoot{The estimated relative uncertainties are $\pm25$\% for $Y_{\rm XPS}$
and $\pm30$\% for $Y_{\rm RAIRS}$.
(*) Nonporous amorphous ice deposited on soot at 120~K and UV
irradiated at 12~K.
($\dagger$) RAIRS-only measurement; the XPS response becomes weakly
sensitive to ice depletion at this thickness.
}
\end{table}

\begin{figure*}[!b]
\centering
\includegraphics[width=13cm]{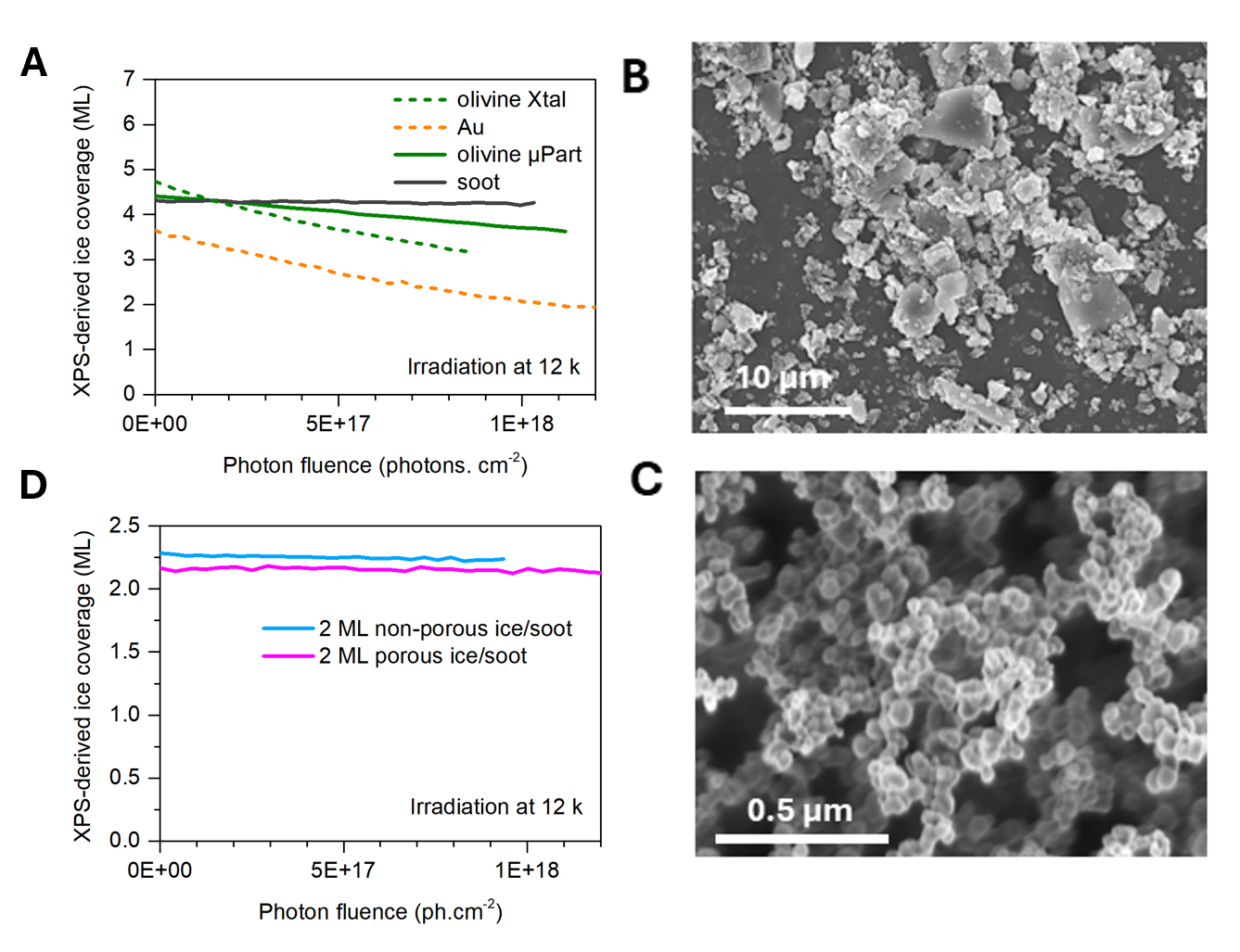}
\caption{Effect of substrate morphology and composition on the UV
photodesorption of water ice at 12~K.
(A) Evolution of the XPS-derived ice coverage as a function of photon
fluence for an initial coverage of approximately 4~ML deposited on
planar gold, planar olivine crystal, olivine microparticle powder, and
soot at 12~K. Linear fits to the coverage curves give photodesorption
yields of $1.5\times10^{-3}$, $1.8\times10^{-3}$,
$7.2\times10^{-4}$, and $7.2\times10^{-5}$~molecule~photon$^{-1}$,
respectively (see Table~\ref{tab:yields}).
(B) Scanning electron microscopy image of the olivine powder film,
showing micrometer- and submicrometer-sized angular particles.
(C) Scanning electron microscopy image of the soot film, showing
highly aggregated carbon nanoparticles with a much finer and more
ramified morphology.
(D) Coverage derived from XPS measurements as a function of UV fluence
at 12~K for 2~ML ice films deposited at 120~K (nonporous ice) and
12~K (porous ice), giving photodesorption yields of
$4.6\times10^{-5}$ and
$1.4\times10^{-5}$~molecule~photon$^{-1}$, respectively.}
\label{fig:morphology}
\end{figure*}
At 12~K, the soot measurements also show an approximately linear coverage dependence over 2--6~ML, but with a much smaller slope than on Au:

\begin{equation}
Y_{\rm XPS}^{\rm soot}(\theta)
\simeq 1.3\times10^{-5}\,\theta
\quad {\rm molecule~photon^{-1}}.
\label{eq:yield_soot}
\end{equation}

Thus, even after normalization by ice coverage, the photodesorption yield on soot at 12~K remains
lower by a factor of about 20--25. To test whether the morphology-induced suppression persists at larger ice coverages, an additional H$_2$O film with a coverage of 10 ML was deposited on soot and irradiated at 12 K. Because the XPS response becomes weakly sensitive to additional ice at such thicknesses, this measurement was quantified from RAIRS only. The photodepletion yield was $3.4\times10^{-4}$ molecule photon$^{-1}$. As observed at lower coverages, the RAIRS-derived yield is somewhat higher than the corresponding XPS trend, but remains within the same low-yield regime characteristic of soot. Importantly, no recovery toward the yields measured on planar substrates is observed, showing that the morphology-induced suppression persists even when the ice coverage approaches the upper range expected for interstellar grain mantles.

\subsection{Effect of substrate composition and particle morphology}
\label{sec:morphology}

To assess the effect of substrate morphology independently of
its composition, additional XPS measurements were performed at 12~K for an
initial ice coverage of approximately 4~ML on planar olivine and an
olivine particle film. The resulting coverage curves are compared with
those obtained on planar gold and soot in Fig.~\ref{fig:morphology}A.
On the two planar substrates, the photodesorption yields are very
similar, with
\(Y_{\rm XPS}=1.5\times10^{-3}\) molecule~photon\(^{-1}\) on gold and
\(1.8\times10^{-3}\) molecule~photon\(^{-1}\) on planar olivine.
In contrast, \(Y_{\rm XPS}\) decreases to
\(7.2\times10^{-4}\) molecule~photon\(^{-1}\) on the olivine particle
film and to \(7.2\times10^{-5}\) molecule~photon\(^{-1}\) on soot.
Thus, for 4~ML of ice, the yield is reduced by a factor of about 2--3
from planar olivine to olivine particles, while on soot it is lower by
a factor of about 10 relative to the olivine particles and by a factor
of about 20 relative to planar gold. The similarity between the two
chemically distinct planar substrates, together with the systematic
decrease on particulate substrates, shows that substrate morphology has
a much stronger influence on photodesorption than the chemical identity
of the underlying material. The resulting sequence of \(Y_{\rm XPS}\)
is therefore

\begin{equation}
Y_{\rm Au}\simeq Y_{\rm olivine\,crystal}
>Y_{\rm olivine\,particles}
>Y_{\rm soot}.
\label{eq:yield_sequence}
\end{equation}

\medskip

The electron microscopy images in Fig.~\ref{fig:morphology}B,C
illustrate the markedly different morphologies of the two particulate
samples. The olivine powder consists predominantly of angular
submicrometer- and micrometer-sized fragments, whereas the soot forms a
much finer and more highly aggregated network of carbon nanoparticles.

\begin{figure}[!htbp]
    \centering
    \includegraphics[width=8cm]{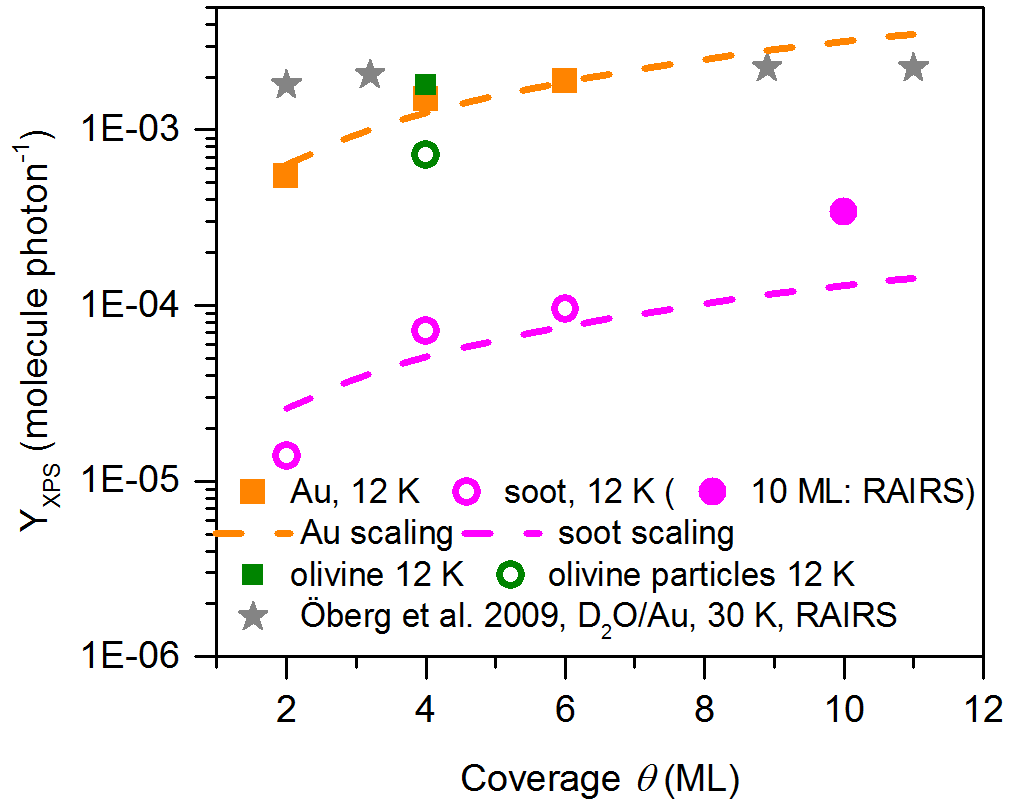}
\caption{Photodesorption yield as a function of initial ice coverage. Symbols show the XPS-derived H$_2$O photodesorption yields at 12 K on planar Au, soot, planar olivine, and olivine particles (note the logarithmic scale). On soot, the 10 ML point was quantified by RAIRS only because the XPS response is weakly sensitive to ice depletion at this thickness. Dashed lines show the
coverage dependence of the photodesorption yield measured for planar
Au and soot, as described by Eqs.~(\ref{eq:yield_au}) and
(\ref{eq:yield_soot}), respectively, and extrapolated to 12~ML here for
illustration. Gray stars show the thickness dependence
reconstructed from the RAIRS data from \citet{Oberg2009}
for D$_2$O on Au at 30~K.}
    \label{fig:yield_coverage}
\end{figure}

Figure~\ref{fig:yield_coverage} summarizes the combined dependence of
the photodesorption yield on ice coverage and substrate morphology. It
shows the linear increase of \(Y_{\rm XPS}\) with ice
coverage observed on both planar Au and soot, as described by Eqs.~(\ref{eq:yield_au}) and
(\ref{eq:yield_soot}) (note the logarithmic scale). The olivine measurements are consistent with the
morphological hierarchy established above. The
D$_2$O/Au thickness dependence from the RAIRS measurements reported by \citet{Oberg2009} is shown for comparison.

\section{Discussion}
\label{sec:discussion}

Collisional retrapping can reduce desorption from porous ice, because
species initially ejected toward the vacuum may collide with pore walls,
lose translational energy, and readsorb before escaping
\citep{Hama2010,Yabushita2006,Kimmel1995,Parent2002,Bertin2012}. To test whether intrinsic ice porosity contributes to the
observed suppression, a water film was deposited on soot at 120~K,
subsequently cooled to 12~K, and irradiated at this temperature. Ice
deposited at 120~K is expected to be compact and essentially
nonporous. Despite this compact structure, its photodesorption remained
strongly suppressed, with
\(Y_{\rm XPS}=4.6\times10^{-5}\) molecule~photon\(^{-1}\), compared
with \(1.4\times10^{-5}\) molecule~photon\(^{-1}\) for a porous film
deposited directly at 12~K (Fig.~\ref{fig:morphology}D). Thus,
intrinsic ice porosity is not the main cause of the observed quenching of
photodesorption.

Geometrical confinement within the nanoscale depressions of the soot surface is unlikely to explain the observed suppression either. These depressions are typically only 0.5 nm deep and 2 nm wide, and should become progressively filled as the ice coverage increases. Yet photodesorption remains strongly suppressed up to 6 ML, with no recovery toward planar-substrate behavior observed in the RAIRS measurements at 10 ML. Simple trapping within these shallow depressions therefore appears insufficient to account for the effect.

Together, these observations suggest that grain morphology may instead act at an earlier stage by modifying the molecular organization of the ice in which photodesorption takes place. Such an interpretation is physically plausible because UV photodesorption in molecular ices depends on the local redistribution of energy and momentum following photon absorption. For instance, in the case of CO, photodesorption proceeds predominantly through an indirect desorption induced by electronic transition (DIET) mechanism involving the outer few molecular layers \citep{Bertin2012}. Recent ab initio molecular-dynamics calculations further show that vibrational relaxation of an excited CO molecule can trigger successive intermolecular energy transfers leading to desorption \citep{DelFre2023}. In H$_2$O ice, the microscopic pathways are different, with UV excitation commonly leading to molecular photodissociation. Molecular-dynamics calculations show that photodesorption is most efficient following excitation in the outer few molecular layers, whereas excitations occurring deeper in the ice increasingly lead to trapping or recombination of the photofragments \citep{Andersson2008,Arasa2010}. For intact H$_2$O desorption, two pathways involving photodissociation have been identified. In the first, H and OH fragments recombine, and part of the released energy drives the desorption of the newly formed water molecule. In the second, an energetic H atom transfers momentum to a neighboring H$_2$O molecule, ejecting it through the kick-out mechanism \citep{Andersson2008,Yabushita2009,Arasa2015}. Desorption of weakly bound adsorbates (Ar, Kr, N$_2$, CO) on water ice driven by this mechanism has also been demonstrated experimentally \citep{Dupuy2021}. A further possibility is exciton-mediated desorption, in which electronic charge redistribution produces repulsive intermolecular forces that may eject an intact H$_2$O molecule without complete photodissociation \citep{DeSimone2013,Fillion2022}. The relative importance of these mechanisms remains uncertain and may depend on the depth of excitation and the local molecular environment, including binding and coordination. Photodesorption thus requires the conversion of absorbed energy into molecular motion sufficient to overcome the binding of a surface molecule. This conversion may be sensitive to the spatial organization of the ice and to energy dissipation at ice–grain interfaces.

An additional indication that photodesorption can be initiated within the ice, rather than exclusively at its outermost surface, is provided by its thickness dependence. In the optically thin regime investigated here, the fraction of absorbed photons increases approximately linearly with ice coverage, and the photodesorption yield measured on planar Au follows the same trend. This behavior is consistent with photoexcitation in subsurface layers contributing to the desorption process.

The structure, morphology and connectivity of the condensed ice may therefore influence the probability that an absorbed photon leads to desorption. On planar substrates, increasing coverage promotes the coalescence of molecular clusters into more continuous films, as observed by scanning-probe microscopy \citep{Mehlhorn2009,Shiotari2017,Koshida2026}. On soot, heterogeneous nucleation and growth may instead preserve a patchy ice distribution over a wider coverage range \citep{Parent2026}, consistent with the persistent XPS substrate signal at high H$_2$O exposures (Fig.~\ref{fig:xps_calibration}B). Ice domains may become several monolayers thick while remaining laterally limited, leaving nearly bare substrate between them. Such a structure would increase the relative importance of domain edges and ice-grain interfaces, where molecular binding and energy redistribution may significantly differ from those in continuous films. The stronger suppression observed on soot than on the coarser olivine particles is consistent with an influence of nanoscale morphology on these processes. Further investigations will be needed to establish how the local ice structure affects the energy-transfer pathways leading to photodesorption and to quantify their contribution to the observed suppression.

\section{Conclusions and astrophysical implications}

Our results show that UV photodesorption of water ice is strongly
influenced by the physical structure of the grain surface. Nanoscale
morphology affects the way the ice grows and is organized, and may
thereby modify the intermolecular energy redistribution that ultimately
leads to desorption. Photodesorption cannot be regarded as a property of
the ice alone, but of the coupled ice--grain system.

These results have two direct implications for astrochemical modeling.

First, laboratory photodesorption yields implemented in astrochemical
models are generally derived from planar substrates and treated as a
single value. Our results show that both simplifications can be
inadequate: photodesorption is substantially reduced on particulate or
aggregated grains and depends on ice coverage in the few-monolayer
regime relevant to interstellar ice mantles, as discussed in Sect.~1.
Photodesorption should therefore be treated as both morphology- and
coverage-dependent in astrochemical models.

Second, nanoscale morphology has previously been shown to reduce
molecular sticking and modify the initial growth of ice
\citep{Laffon2021,Parent2026}, thereby slowing molecular freeze-out and
ice accumulation \citep{Stadler2025}. Taken together, these results
indicate that grain morphology affects both accretion and loss
processes, with opposite consequences for the gas--ice partitioning.
Although fewer H$_2$O molecules and photoproducts escape from
particulate grains, those reaching the gas phase also recondense less
efficiently because of the lower sticking coefficients. Photodesorption
and sticking should therefore be treated jointly in astrochemical
models, rather than adjusting the photodesorption yield alone.

\begin{acknowledgements}
The authors thank G. Arthaud for technical support, O. Grauby for the SEM imaging, D. Ferry for the TEM measurements used to characterize the nanoscale surface roughness, and F. Berger for glassblowing the UV lamp.
\end{acknowledgements}

\bibliographystyle{aa}

\bibliography{sources}

\
\clearpage

\pagebreak

\clearpage

   \FloatBarrier

\begin{appendix}
\nolinenumbers

\section{Estimate of the ice coverage of interstellar grains}
\label{app:coverage}

We use the Mathis--Rumpl--Nordsieck (MRN) grain-size distribution for graphite and olivine \citep{Mathis1977}, which can be written as

\begin{equation}
\log n_i(a)=K_i-3.5\log(a),
\label{eq:mrn}
\end{equation}

where $n_i(a)$ is the number of grains of component $i$ per H nucleus and per $\mu$m of grain radius, with $K_{\rm C}=-15.24$ for graphite and $K_{\rm Ol}=-15.21$ for olivine. This analytical representation was shown by MRN to reproduce satisfactorily the interstellar extinction over $0.005<a<0.25$~$\mu$m. Following \citet{Hollenbach2009}, we extend this distribution down to $a_{\rm min}=20$~\AA\ ($0.002$~$\mu$m), below which transient photon heating prevents the stable accumulation of ice, and adopt $a_{\rm max}=0.25$~$\mu$m for the ice-bearing grain population.

The total geometrical cross section of the grains per H nucleus is then

\begin{equation}
\sigma_{\rm H}
=
\int_{a_{\rm min}}^{a_{\rm max}}
\pi a^2 n_{\rm tot}(a)\,{\rm d}a,
\label{eq:sigma_int}
\end{equation}

where

\begin{equation}
n_{\rm tot}(a)
=
\left(10^{K_{\rm C}}+10^{K_{\rm Ol}}\right)a^{-3.5}.
\end{equation}

Using $a$ in $\mu$m and converting the resulting area to cm$^2$,

\begin{equation}
\sigma_{\rm H}
=
2\pi\times10^{-8}
\left(10^{K_{\rm C}}+10^{K_{\rm Ol}}\right)
\left(
a_{\rm min}^{-1/2}-a_{\rm max}^{-1/2}
\right),
\label{eq:sigma}
\end{equation}

which gives

\begin{equation}
\sigma_{\rm H}\simeq1.5\times10^{-21}
~{\rm cm^2\,H^{-1}}.
\label{eq:sigma_value}
\end{equation}

This value is close to the approximate $2\times10^{-21}$~cm$^2$~H$^{-1}$ adopted by \citet{Hollenbach2009} for an MRN distribution with an effective lower cutoff of 20~\AA. They also emphasize that this cross section is uncertain, in particular because coagulation can reduce the available grain area whereas ice growth can increase it.

For spherical grains, the physical surface area is four times their geometrical cross section. The total surface area available for ice per H nucleus is therefore

\begin{equation}
A_{\rm surf/H}
=
4\sigma_{\rm H}
\simeq6.1\times10^{-21}
~{\rm cm^2\,H^{-1}}.
\label{eq:surface}
\end{equation}

Taking a surface density of adsorption sites $N_{\rm s}\simeq10^{15}$~cm$^{-2}$, as commonly adopted for water ice, one complete monolayer over this grain population corresponds to an abundance

\begin{equation}
X_{\rm 1ML}
=
A_{\rm surf/H}N_{\rm s}
\simeq6.1\times10^{-6}
\label{eq:x1ml}
\end{equation}

H$_2$O molecules per H nucleus.

The mean effective number of water-ice monolayers corresponding to an abundance measured along the line of sight is consequently

\begin{equation}
N_{\rm ML}
=
\frac{N({\rm H_2O})/N_{\rm H}}
{X_{\rm 1ML}},
\label{eq:nml}
\end{equation}

where $N_{\rm H}=N({\rm H\,I})+2N({\rm H_2})$ is the total hydrogen column density. \citet{Boogert2015} compiled H$_2$O ice abundances toward background stars probing quiescent molecular clouds and dense cores. Their distribution has a median abundance of $40\times10^{-6}$ relative to H nuclei, with the central 50\% of the observed values between approximately $31\times10^{-6}$ and $48\times10^{-6}$. Equation~\ref{eq:nml} therefore gives a median effective coverage of

\begin{equation}
N_{\rm ML}\simeq6.6~{\rm ML},
\end{equation}

with the central 50\% of the observations corresponding to approximately

\begin{equation}
5\text{--}8~{\rm ML}.
\end{equation}

The full observed abundance range, from $>9\times10^{-6}$ to $62\times10^{-6}$, corresponds to approximately $>1.5$ to 10~ML. Consistently, \citet{Boogert2015} note that the lowest detectable H$_2$O ice columns, around $10^{17}$~cm$^{-2}$, correspond to only a few monolayers for an MRN grain population.

An independent constraint follows from the elemental oxygen budget. \citet{Boogert2015} estimate an upper mantle thickness of approximately 0.005~$\mu$m (50~\AA) for H$_2$O-rich mantles under an MRN distribution. For compact amorphous water ice, using a molecular density corresponding to $\rho\simeq0.94$~g~cm$^{-3}$ and $N_{\rm s}\simeq10^{15}$~cm$^{-2}$, one monolayer has an effective thickness of about 3.2~\AA, so that 50~\AA\ corresponds to approximately 15~ML.

\section{XPS determination of water-ice coverage and yield uncertainties}
\label{app:xps}

\begin{figure}
\centering
\includegraphics[width=\columnwidth]{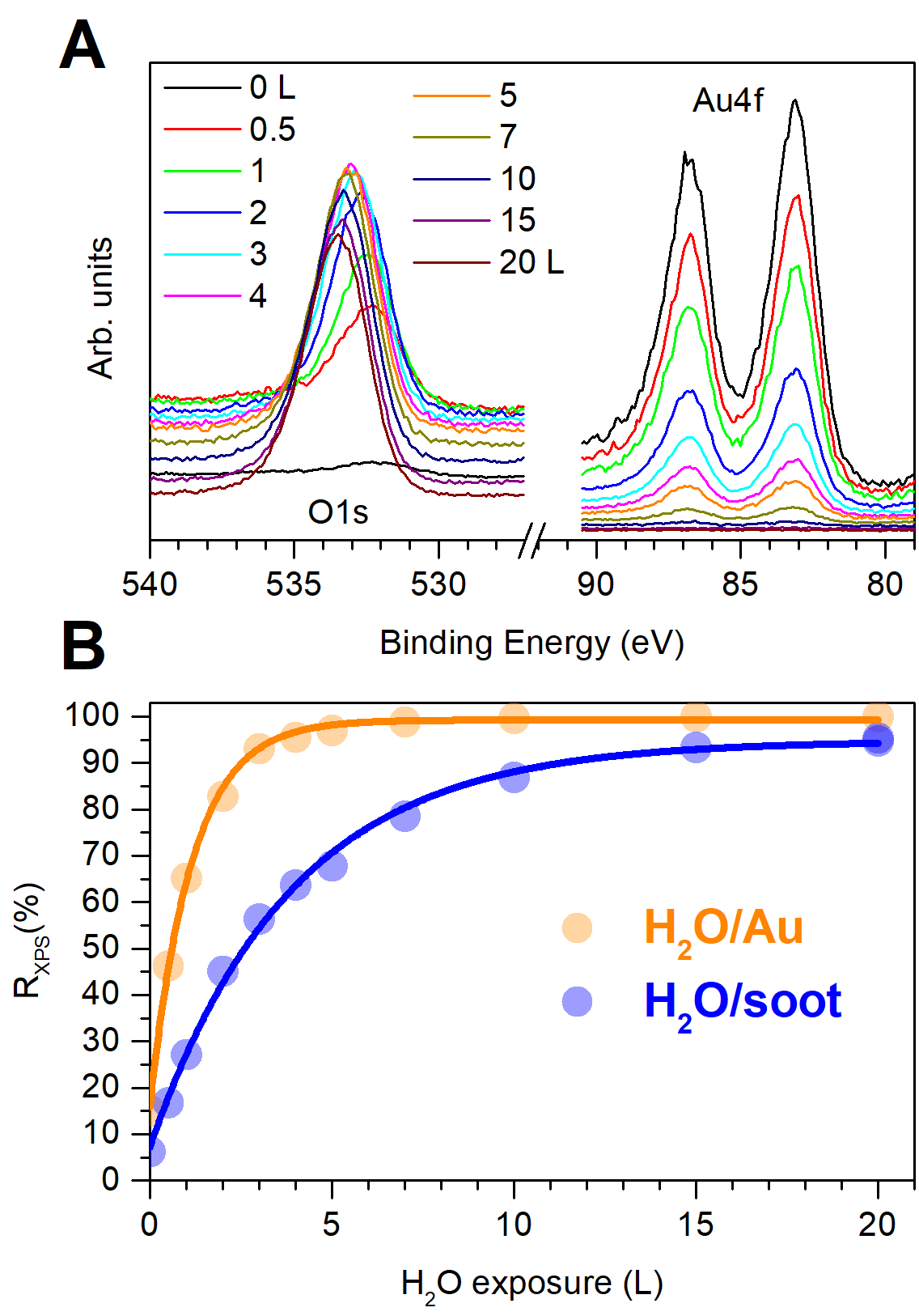}
\caption{XPS calibration of the H$_2$O coverage on gold and soot.
(A) Evolution of the O~1s adsorbate signal and attenuation of the Au~4f substrate signal during H$_2$O deposition on gold at 12~K. The indicated values correspond to the incident H$_2$O exposures in Langmuir (L). The same procedure was applied to soot using the C~1s signal as the substrate reference.
(B) RSF-corrected XPS adsorbate fraction, $R_{\rm XPS}$, expressed in percent and plotted as a function of incident H$_2$O exposure for gold and soot. Symbols represent the experimental data and solid lines the exponential fits. Comparison of the exposures required to reach the same $R_{\rm XPS}$ on both substrates was used to determine the H$_2$O sticking coefficient on soot relative to gold. Analytical inversion of the fitted curves was subsequently used to convert the XPS signal measured during UV irradiation into an instantaneous water coverage.}
\label{fig:xps_calibration}
\end{figure}

Figure~\ref{fig:xps_calibration} illustrates the procedure used to convert the XPS signals into water-ice coverages. Fig. ~\ref{fig:xps_calibration}A shows the increase in the H$_2$O O~1s signal near 533~eV (left) and the simultaneous attenuation of the Au~4f substrate signal (right) during water deposition (1~L $=10^{-6}$~Torr~s) on gold at 12~K. The same procedure was applied to soot and olivine using the substrate emission lines as the substrate reference.

After background subtraction, the integrated XPS peak areas $A$ were corrected using their respective relative sensitivity factors (RSFs). The normalized XPS adsorbate fraction $R_{\rm XPS}$ was calculated as

\begin{equation}
R_{\rm XPS}
=
\frac{A_{\rm O\,1s}/{\rm RSF}_{\rm O\,1s}}
{
A_{\rm O\,1s}/{\rm RSF}_{\rm O\,1s}
+
A_{\rm sub}/{\rm RSF}_{\rm sub}
}.
\label{eq:rxps}
\end{equation}

Here, $A_{\rm sub}$ corresponds to the Au~4f signal for gold, to the C~1s signal for soot, and to the Mg~2p + Si~2p signals for olivine. Figure~\ref{fig:xps_calibration}B shows $R_{\rm XPS}$ as a function of H$_2$O exposure for the Au and soot substrates. This normalization procedure also accounts for differences in effective surface area and for geometrical shadowing of photoelectrons emitted within the particulate film \citep{Laffon2021}.

These uptake curves were first used to determine the H$_2$O sticking coefficient on the different substrates. A preliminary RAIRS calibration on gold showed that 1~ML of water is condensed per Langmuir of exposure. The H$_2$O sticking coefficient on gold, $S_{\rm Au}$, was therefore taken as unity, and the gold uptake curve provides the reference response for unit sticking. For soot, the relative sticking coefficient was determined by comparing the incident exposures required to reach the same value of $R_{\rm XPS}$:

\begin{equation}
S_{\rm soot}
=
\frac{E_{\rm Au}}{E_{\rm soot}},
\label{eq:ssoot}
\end{equation}

where $E_{\rm Au}$ and $E_{\rm soot}$ are the incident exposures required to reach the same normalized adsorbate fraction on gold and soot, respectively. Comparison over the common range of the two uptake curves yielded

\begin{equation}
S_{\rm soot}\simeq0.2.
\end{equation}

The same procedure was applied to each substrate, for which the quantity $S\times E$ provides the effective coverage at any H$_2$O exposure (in ML), as quoted in Figs.~\ref{fig:gold_soot} and \ref{fig:morphology}.

Under UV irradiation, the initial ice coverage is expected to decrease with irradiation time. We can use the coverage curves, like those presented in Fig.~\ref{fig:xps_calibration}B, to assess how much of the ice thickness is lost upon irradiation. To this end, the uptake curve obtained for each substrate was fitted with an exponential function,

\begin{equation}
R_{\rm XPS}(E)
=
R_{\infty}
-
\left(R_{\infty}-R_0\right)\exp(-kE),
\label{eq:uptake}
\end{equation}

where $R_0$ and $R_{\infty}$ are the initial offset and asymptotic adsorbate fractions, respectively, and $k$ is the exponential rise constant. Analytical inversion gives

\begin{equation}
E_{\rm eq}
=
-\frac{1}{k}
\ln
\left(
\frac{R_{\infty}-R_{\rm XPS}}
{R_{\infty}-R_0}
\right).
\label{eq:inverse_uptake}
\end{equation}

Each value of $R_{\rm XPS}$ measured during UV irradiation could thus be converted into an equivalent exposure and subsequently into an instantaneous water coverage.

As the ice thickness approaches the effective attenuation length of the O~1s photoelectrons, the uptake curves progressively approach $R_{\infty}$, reducing their sensitivity to further changes in condensed-phase coverage. The experiments were therefore restricted to initial exposures of approximately 6~L on gold, corresponding to approximately 6~ML, and 30~L on soot, corresponding to approximately $30\times0.2=6$~ML equivalent.

\medskip
The uncertainty on the XPS-derived photodesorption yields $Y_{\rm XPS}$ results mainly from the conversion of the XPS signal into ice coverage, the determination of the depletion slope, and the calibration of the incident UV photon flux. Based on the error analysis established for the same XPS methodology \citep{Laffon2021}, the uncertainty associated with the coverage determination is estimated to be about 10\%. The UV photon flux is known within approximately $\pm10$\%, while the uncertainty associated with the fitted slopes is conservatively estimated to be below about $\pm20$\%. Assuming these contributions to be independent and adding them in quadrature gives an overall uncertainty of approximately $\pm25$\% on $Y_{\rm XPS}$.
For the RAIRS-derived yields $Y_{\rm RAIRS}$ on gold, the main uncertainties arise from the conversion of the integrated OH-stretching band into ice coverage, the fitted depletion slope, and the incident UV photon flux. \citet{Bouilloud2015} estimated an uncertainty of about 20\% associated with integration boundaries
and baseline subtraction in infrared band measurements. Combining this
contribution with the approximately $\pm10$\% uncertainty on the UV photon flux and a conservative $\pm20$\% uncertainty on the fitted slopes gives an estimated relative uncertainty of about $\pm30$\% on $Y_{\rm RAIRS}$ for gold. For the RAIRS-derived yields on soot, the infrared signal was scaled to the initial equivalent coverage determined from the XPS-based uptake calibration. In addition to the uncertainties associated with the infrared signal, fitted depletion slope, and photon flux, this procedure introduces an uncertainty in the initial coverage, estimated to be about 10\%. Combining these contributions gives an estimated relative uncertainty of approximately ±30\% on \(Y_{\rm RAIRS}\), comparable to that on gold.

\end{appendix}
\end{document}